# Optimisation and Loss Analyses of Pulsed Field Magnetisation in a Superconducting Motor with Cryocooled Iron Cores

Qi Wang, Luning Hao, Hongye Zhang, Guojin Sun*, Haigening Wei, Yuyang Wu, Zhipeng Huang, Jintao Hu and Tim Coombs

Q. Wang, L. Hao, H. Wei, Z. Huang, Y. Wu, J. Hu, T. Coombs are with the Electrical Engineering Division, Department of Engineering, University of Cambridge, Cambridge CB3 0FA, U.K.
Hongye Zhang is with the Institute for Energy Systems, School of Engineering, The University of Edinburgh, Edinburgh EH9 3FB, UK.
Guojin Sun is with the College of Engineering, Qinghai University of Technology, Qinghai 810016, China.
*Corresponding author.

*Abstract*— A 2D electromagnetic-thermal coupled numerical model has been developed using the finite element method and validated against experimental data to investigate a superconducting machine featuring high-temperature superconducting (HTS) tape stacks and cryocooled iron cores. The HTS stacks are transformed into trapped field stacks (TFSs) through pulsed field magnetisation (PFM), generating rotor fields. After PFM, the superconducting motor operates on the same principle as permanent magnet synchronous motors. This study explores the behaviour of HTS stacks by altering the stack's layer number from one to nine and adjusting the pulsed current amplitude from 250 A to 1000 A. The primary objective of this paper is to identify the optimal combination of pulsed current amplitudes and TFS layer numbers for achieving maximum magnetisation fields. The secondary objective is to evaluate the overall losses in both superconducting and non-superconducting parts of the machine during magnetisation, including heat generated in various layers of the TFS, and losses in the motor's active materials (copper windings and iron cores). Two motor configurations were proposed, and two calculation methods using linear interpolation of iron losses and steel grades were introduced to estimate the iron losses for the studied iron material, M270-35A. This pioneering study is expected to serve as a valuable reference for loss analyses and structural design considerations in developing superconducting machines.

## Contents

## 1. Introduction

The surge in interest in electric aircraft arises from mounting apprehensions about energy scarcity and environmental considerations [1] [2] [3]. Superconducting motors incorporating high-temperature superconducting (HTS) materials have emerged as a compelling choice for effective propulsion solutions thanks to their advantages, such as reduced dimensions, minimal energy losses, and increased power density [4] [5].

In addition to coil-shaped HTS magnets [6] [7], both superconducting bulks and stacked HTS coated conductors (CCs) are able to trap magnetic flux upon magnetisation, transforming into trapped field magnets (TFMs). Stacked HTS CCs and HTS bulk samples with trapped fields refer to trapped field stacks (TFSs) and trapped field bulks (TFBs), respectively. As trapped field magnets (TFMs), stacked HTS CCs offer superior qualities over HTS bulks due to their enhanced mechanical and thermal properties [8], shape flexibility [9], homogenous superconducting properties [10], and superior performance under cross field demagnetisation [11]. Recently, TFMs have emerged as potential replacements for traditional permanent magnets (PMs) [12] [13], particularly in applications such as generating rotor fields for motor operation, notably in synchronous machines [14] [15]. In synchronous machines, conventional rotors utilise either wound field coils or PMs to generate the magnetic field. However, wound field coils face challenges due to the utilisation of brushes and slip rings [16], as well as issues associated with continual current sources typically supplied by current leads [16]. In contrast, PMs, which can be substituted with TFBs [17], or TFSs [18] for superconducting motors, offer more compact designs and diminished thermal and electrical losses caused by power supplies and current leads, thereby leading to higher power densities.

Several methods are commonly utilised for the magnetisation of HTS conductors intended for use as TFMs in superconducting machines, including field cooling (FC) [13], zero field cooling (ZFC) [19], pulsed field magnetisation (PFM) [12] [20], and flux pumping [21] [22]. Among these, the portable, cost-effective, and compact *in-situ* magnetisation process, PFM, is favored for electrical machine applications [19]. However, during PFM, the resultant heat impedes the attainment of the trapped field flux potential in CCs [19]. Therefore, various strategies have been proposed to tackle the issue of heat generation, aiming to improve trapped fields and power densities. These include employing multi-pulse techniques [23], modifying the configuration of magnetisation coils [24], incorporating ferromagnetic/superconducting structures [25], leveraging flux jumps [26], and implementing active waveform control [27].

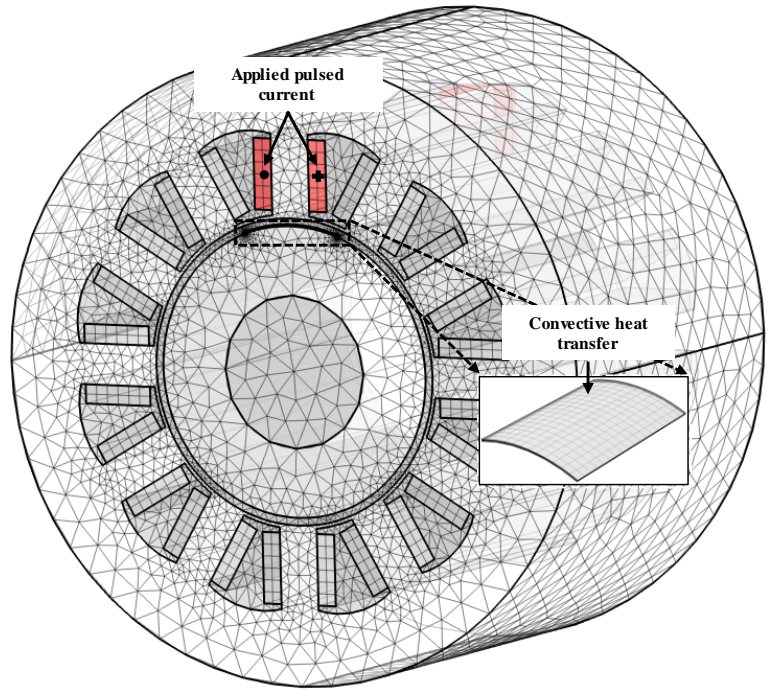

Figure 1. Setup of the studied machine with a TFS mounted on the rotor surface for the pulsed field magnetisation

To investigate the electromagnetic and thermal characteristics of TFSs in HTS motors, a 2D electromagnetic-thermal coupled model was developed using the finite element method (FEM) within a superconducting machine [28], as shown in Figure 1. The machine prototype under examination is a synchronous machine adapted from a surface-mounted permanent magnet motor, incorporating a fractional slot concentrated winding (FSCW) configuration. This HTS machine is constructed based on a conventional machine architecture, featuring iron yokes in both the stator and rotor. In contrast to conventional iron-cored machines, air-cored machines have the potential to achieve higher magnetic loading and fully exploit the capabilities of HTS CCs [29].

However, they are vulnerable to increased air gap harmonics originating from the stator, which can result in a more significant decay of the trapped fields of TFSs. Conversely, conventional iron-cored machines encounter limitations in their magnetic loading capacity due to iron saturation levels [30].

This paper builds on our previous work [28], where the adapted modelling methodology has been validated against experimental data. In this research, two significant issues related to PFM in superconducting machines are addressed, which have not been previously examined for TFSs. The first issue concerns identifying the optimal combination of pulsed current amplitudes and layer numbers of TFSs for PFM to achieve the maximum magnetisation fields. A lower magnetisation field necessitates a small pulsed current but may not fully penetrate stacked HTS tape-magnets. Conversely, excessively large fields result in increased heat generation within TFSs, raising temperatures and thereby reducing trapped field strength. Moreover, a smaller number of HTS tapes requires a lower magnetisation field, reducing the demand for magnetisation but potentially trapping an insufficiently small field, thereby diminishing machine output power. On the other hand, a large number of HTS tapes increases the requirement for magnetisation field strength and further requires a more powerful magnetisation fixture. Consequently, there exists an optimal magnetisation field for a given number of HTS tape layers. Conversely, for a specified pulsed current amplitude, there exists an optimal number of layers to achieve maximum magnetic flux density trapping. Consequently, this study partly focuses on investigating how the trapped field of a single TFS varies with different layer numbers and pulsed current magnitudes during PFM in the examined HTS motor, aiming to present an approach to determine the optimal layer number for achieving the maximum trapped field in an HTS motor.

The second issue involves evaluating the overall losses in both the superconducting and non-superconducting parts of the machine during magnetisation. For the superconducting part, this includes the magnetisation losses due to heat generation in the TFS. For the non-superconducting parts, it encompasses the copper losses in the magnetisation windings in the stator and the iron losses in the stator and rotor iron yokes. At cryogenic temperatures (CT), heat generated in the machine's active materials can be conducted away more efficiently than in the air. However, it is still crucial to thoroughly analyse and reduce the generated losses through proper measures, as these losses can lead to reduced efficiency and shortened machine lifespan. Consequently, for electric motors that incorporate both iron cores and HTS materials, it is essential to investigate the behaviour of iron cores at CT, specifically the iron losses. In this paper, a linear interpolation approach based on steel grades has been developed for the iron material utilised in the HTS machine. This method enables the estimation of iron losses with reasonable accuracy, thereby avoiding the need for complex measurements.

Specifically, this paper comprehensively analyses three elements: one TFS and the two active materials of the machine - iron and copper, to provide a thorough investigation of the magnetisation model in a superconducting motor with cryocooled iron cores. The analysis includes an examination of the influence of the magnetisation field and layer number of the stack on HTS losses. Moreover, the impact of temperature on the electrical conductivity of copper and iron is considered for copper and iron losses. Additionally, two design concepts, distinguished by a warm stator and a cold stator, have been analysed for iron losses.

## 2. Modelling methodology

For the modelling of PFM for HTS stacks, a FEM-based electromagnetic-thermal coupled model [28] was constructed in the commercial software COMSOL Multiphysics 6.1 and validated against the experimental results measured in [31]. For the electromagnetic modelling, the $\boldsymbol{H}$-formulation [32] was used for the whole machine following the governing equations:

$$\mu \frac{\partial \boldsymbol{H}}{\partial t} + \nabla \times \rho(\nabla \times \boldsymbol{H}) = 0 \tag{1}$$

$$E = E_0 (\frac{\boldsymbol{J}}{J_c(\boldsymbol{B},T)})^{n(\boldsymbol{B})} \tag{2}$$

where $\mu$ is the magnetic permeability, $\rho$ is the resistivity of the materials, $E_0$ presents the characteristic electric field with $E_0 = 10^{-4} \text{ V} \cdot \text{m}^{-1}$, and the field-dependent exponent $n(\boldsymbol{B})$ was taken from [13]. The anisotropic critical current density employed is dependent on the magnetic field and temperature, and it can be expressed by the product of the magnetic measured field-dependent current density $J_c(\boldsymbol{B})$ from [8] and the temperature dependence coefficient $\alpha(T)$ ranging between 0 and 1, taken from [33]:

$$J_c(\boldsymbol{B},T) = J_c(\boldsymbol{B}) \cdot \alpha(T) \tag{3}$$

Given the non-negligible heat generation during PFM, a thermal model was constructed to incorporate heat transfer within the PFM process and interconnected with the electromagnetic model [34]. The coupling of both models is bi-directional. The heat transfer model offers real-time temperature inputs to the electromagnetic model, while the electromagnetic model provides the current density $\boldsymbol{J}$ and electrical field $\boldsymbol{E}$ of the stack. The governing equation for the heat transfer module:

$$\rho_{\mathrm{m}} C_p \frac{\partial T}{\partial t} - \nabla \cdot k \nabla T = Q(T) \tag{4}$$

where $\rho_{\mathrm{m}}$ is the mass density, $C_p$ denotes the specific heat capacity, $k$ is the thermal conductivity, and $Q$ is the volumetric heat source for the thermal model, and equals $\boldsymbol{E} \cdot \boldsymbol{J}$. The heat capacity, and thermal conductivity of the CCs are from [35].

The magnetisation of a single superconducting stack in the machine was conducted in the developed FEM model. Figure 1 illustrates the coupled electromagnetic and heat transfer models in the simulated machine. The geometries of the motor and tape stacks are the same as in [8] and [12]. The motor has a stator with an external diameter of 236 mm and an internal diameter of 120 mm, a machine length of 160 mm, and a 3 mm long airgap. Every layer of the AMSC tape was modelled to demonstrate the effect of different layers during PFM. The thickness of one single tape is 87 μm and the layer number varies from one to nine. Due to the symmetry of the machine model, only one-eighth of the machine geometry was utilised for the simulation, where half of the HTS stack was contained. As per [8], the employed AMSC tapes are Rolling-Assisted Biaxially Textured Substrate (RABiTS) YBCO CCs with magnetic substrates Ni-5at.%W.

The heat transfer module was exclusively applied within the HTS stack domain, owing to the transient nature of PFM and its brief timespan. Therefore, only heat generated within the stack was taken into consideration, same as in [36]. Futhermore, a convective heat transfer approach employing a heat transfer coefficient of $100\,\mathrm{W}\cdot(\mathrm{m}^2\cdot\mathrm{K})^{-1}$ was adopted. In the original experimental setup in [31], the HTS CCs were arranged in a loose stack using Kapton tapes and secured on the rotor surface by a G10 sleeve situated in the airgap. This configuration facilitated full immersion of the HTS stack in liquid nitrogen. As a result, the convective heat flux boundaries are now applied to the perimeters of the HTS stack in the current simulation model, as depicted in Figure 1. The ambient temperature was established at the boiling point of liquid nitrogen, equivalent to 77 K. The stator and rotor yokes were constructed from silicon iron M270-35A, with its *B-H* curve at 77 K (converted using the relative permeability at 77 K obtained in section III) integrated into the magnetisation model for the iron region. The remaining components of the machine were modelled as air.

Table 1. The magnetic field magnitude in the airgap 1 mm above the stack centre point

| Pulsed current magnitude [A] | Airgap field [T] |
|---|---|
| 250 | 0.88 |
| 500 | 1.10 |
| 750 | 1.28 |
| 1000 | 1.43 |

For the PFM, the waveform of the applied current pulse was obtained from [31]. The magnitude of the current pulse has been adjusted proportionally to the applied current amplitude. The pulsed current for the PFM in this study ranges from 250 A to 1000 A, with increments of 250 A. Additionally, the layer number within the stack was varied between one to nine inclusively. Prior to implementing PFM for magnetising the tape stack, a series of simulations were executed utilising the machine model without the tape stack, to ascertain the amplitude of the pulsed field applied to the magnetisation in the air gap. The results are shown in Table 1.

## 3. Calculation of copper loss and iron loss

The current machine model is based on experimental data from [31], where the entire machine was submerged in liquid nitrogen to cool the HTS CCs. The simulation for the trapped field during PFM is based on this experimental setup. However, this approach is inefficient due to the large heat capacity of the iron components, resulting in lengthy cooling times. In the configuration explored in this paper, which involves a partially superconducting motor with only the rotor being superconductive, a more efficient design option entails cooling the rotor to 77 K while maintaining the stator, including the stator iron yoke and windings, at RT. This concept can be referred to as the 'cold rotor' (CR) concept, while the former can be termed the 'cold motor' (CM) concept. While the CM concept is straightforward, the process of cooling down and warming up the entire machine takes days.

In contrast, the CR concept reduces the cooling time and cryogenic system costs. However, this approach requires a robust cooling system and sealing measures to prevent leaks from the rotating rotor containing liquid nitrogen during motor operation. Consequently, the temperature of different parts of the machine under the two concepts can be summarised in Table 2. For loss calculations in the current machine model during PFM, both concepts will be considered, with the CM concept being the realistic one and the CR concept being an ideal one. Additionally, it is important to note that the parameters of the main construction materials of the machine except superconductors, e.g., copper and iron, vary at different temperatures (293 K and 77 K).

Table 2. Temperature of different parts in the machine for the CM and CR concept

| Concept | Stator | Rotor |
|---|---|---|
| CM | 77 K | 77 K |
| CR | 293 K | 77 K |

When computing losses in copper and iron, direct comparison with typical values or measured loss data from other machines is not feasible. Therefore, specific losses, obtained by dividing the absolute losses by the weights of the active materials, provide more reliable references. Table 3 lists the mass of the active material parts of the machine, including the stator copper windings employed for magnetisation (marked red in Figure 1), the stator iron, and the rotor iron.

Table 3. Mass of the active material parts

| Active material part | Value [kg] |
|---|---|
| Stator copper windings for magnetisation | 0.25 |
| Stator iron | 32 |
| Rotor iron | 9 |

### 3.1. Copper loss

Due to the copper windings in the stator of the machine, copper losses occur in the coils into which the pulsed current is injected. Solid copper bars are used as stator windings in the machine. According to Ohm's law, these losses depend solely on the magnitude of the pulsed current and the resistance of the stator windings, which serve as the magnetisation windings:

$$P_{Cu} = L \cdot \frac{1}{T} \int_0^T J \cdot E dt \tag{5}$$

where $L$ represent the length of the stator winding in the motor's axial direction, $J$ and $E$ stand for the current density, and the electric field within the magnetisation stator windings, and $T$ denotes the duration of the magnetising pulse.

### 3.2. Iron loss

In this section, the employed Bertotti iron loss model and the methodology to calculate the iron losses in the machine model are presented. For accurate measurement of iron losses, the experimental devices and control setup should comply with the conditions specified in DIN 60404-6 [37]. A linear interpolation approach based on steel grades has been devised in this research for the iron material utilised in the HTS machine, specifically M270-35A. This approach provides an acceptable estimation of the loss data of M270-35A at 77 K without complicated measurements at low temperatures.

During the magnetisation process, the diamagnetism exhibited by the superconducting materials could potentially shield the magnetic field within the rotor iron, thereby influencing the losses in both the stator and rotor iron yokes. This phenomenon is not commonly observed in conventional permanent magnet synchronous motors, of which the magnetisation process does not occur within the motors themselves. To examine and analyse the influence of the magnetisation current magnitude on individual and total iron losses, an iron loss model was established for the machine without HTS materials. Additionally, the impact of HTS stacks on the iron losses can be assessed by incrementally increasing the number of layers in the HTS stack.

#### 3.2.1. Iron loss model

Three types of iron losses are typically considered for the iron loss calculation: hysteresis loss, eddy-current loss, and excess loss [38]. The Bertotti equation [39] is utilised for this calculation and is expressed as follows:

$$P_{Fe} = K_h \cdot f \cdot B^\alpha + K_c \cdot f^2 \cdot B^2 + K_e \cdot f^{1.5} \cdot B^{1.5} \tag{6}$$

where $\alpha$ is the exponent for the magnetic flux density for hysteresis loss, $K_h$, $K_c$, and $K_e$ are the coefficients for the hysteresis loss, the eddy current loss, and the excess loss, respectively. The exponent $\alpha$, and coefficients $K_h$, $K_e$ can be determined through curve fitting with the software ANSYS Motor-CAD based on iron loss data, while $K_c$ is determined through the following equation [38]:

$$K_c = \frac{\pi^2 d^2 \sigma}{6} \tag{7}$$

where $\sigma$ is the electrical conductivity of M270-35A and $d$ represents the thickness of the laminated iron sheet.

The total iron loss in the entire motor consists of the iron losses occurring in the stator and rotor, respectively, as described in Equation (8). The losses in both the stator and rotor iron yokes comprise the same components: hysteresis, eddy current, and excess losses. Therefore, the overall iron losses consist of six components, which are the three types of losses in the stator and rotor iron yokes, respectively.

$$Q_{Fe,total} = Q_{Fe,stator} + Q_{Fe,rotor} = \int_0^\tau \int_{V\_stator} P_{Fe,stator} dV dt + \int_0^\tau \int_{V\_rotor} P_{Fe,rotor} dV dt \tag{8}$$

The *B-H* curve is a crucial parameter for determining the iron losses in a given material. For M270-35A, it is assumed that its *B-H* curve remains constant at both RT and 77K. This assumption is based on the observation that the *B-H* curves of three other similar steel grades (M330-50A, M400-50A, and M530-50A) exhibit negligible difference between RT and 77K [40]. The *B-H* of M270-35A at RT can be found in [41].

Frequency is another critical parameter for the iron loss calculation. The frequency in Equation (6) typically represents the frequency of the AC field in the iron. However, for a single magnetic field pulse, there are abundant frequency components. Therefore, a fast Fourier transform (FFT) is conducted to identify the most significant harmonics, and the frequencies of these harmonics are utilised in Equation (6). Figure 2 (a) demonstrates the shape of the pulse utilised for the PFM, with a time length of $\tau = 11$ μs and the magnitude scaled to one. The FFT decomposition of the pulse is illustrated in Figure 2 (b), showing that the most significant harmonics of the FFT are of the first and the second orders. The zeroth order corresponds to the DC offset value of the pulse and does not contribute to the AC losses in the iron yokes. Therefore, the frequencies of the first two orders (84.7 Hz and 169.4 Hz) are employed for calculating the iron loss, with their respective amplitude proportions relative to the DC component. For the simulation in COMSOL, a Time to Frequency Losses solver is employed to calculate the stator iron losses with the Bertotti equation. By integrating the volume-based iron losses in W/m$^3$ with the stator iron volume, the iron losses power in W can be obtained, which will be demonstrated in the results section 4.2.

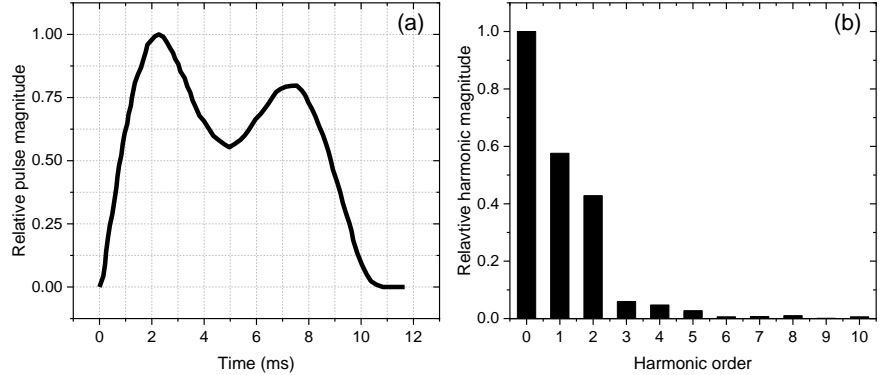

Figure 2. The employed pulse with magnitude scale to one (a) shape (b) the FFT decomposition

### 3.2.2. Calculation methodology

Due to the absence of measured loss data for the iron material M270-35A at 77 K, directly calculating the iron loss at this temperature is not feasible. Thus, the first step is to devise an approach for obtaining the iron loss data at 77 K for M270-35A. Fortunately, measurement data are available for two similar materials M235-35A and M330-35A, which were tested at both RT and 77 K at various frequencies and magnetic flux densities [42].

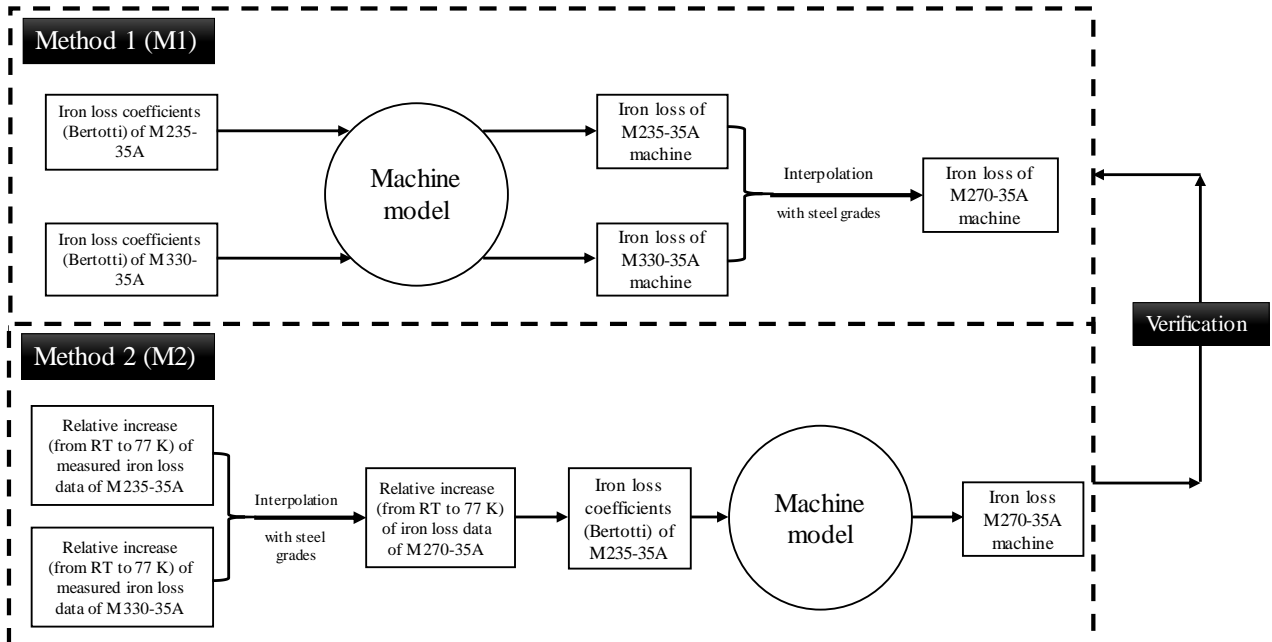

Figure 3. Methodology for calculating the iron loss of M270-35A machine with the available iron loss parameters of M235-35A and M330-35A

All three materials belong to the non-oriented (NO) electrical steels on the powercore® standard grades manufactured by Thyssenkrupp [43]. In the material label, the designation '35A' denotes the lamination sheet thickness of 0.35mm while the steel grades 235, 270, and 330 represent the maximum core loss at 50 Hz, which are 2.35, 2.70, and 3.30 W·kg$^{-1}$, respectively. Consequently, it is intuitive to conclude that the material loss characteristics of M270-35A lie between those of M235-35A and M330-35A. The correlation between iron losses and steel grades can be substantiated by the loss data from three different iron materials: M330-50A, M400-50A, and M530-50A [40]. Notably, the higher-grade M530-50A demonstrates the highest iron losses at both RT and CT, whereas the lower-grade M330-50A exhibits the lowest iron losses. Accordingly, it is assumed that a linear relationship exists between the iron losses and the grades of the steels.

Two methods (Method 1 and Method 2, abbreviated as M1 and M2, respectively) have been employed for the iron loss calculation, as shown in Figure 3. The first method involves direct interpolation between the iron losses of M235 machine (with iron cores made from M235-35A) and M330-35A machine. The second method utilises an indirect approach by obtaining the iron loss data of M270-35A through relative increase from RT to 77 K. The second approach is employed to verify the calculation results computed with the first one.

### 3.2.2.1. Method 1 (M1) – linearity of loss

Method 1 estimates the iron losses of M270-35A by employing linear interpolation, predicted on the assumption that a consistent linear correlation exists between iron loss and steel grade for identical lamination thickness. This approach is expected to be acceptable when the variation in steel grades is limited, which in our instance ranges from 235 to 330. The linearity of the loss data can be presented as follows, by denoting the loss data of M235-35A, M270-35A, and M330-35A as $\alpha$, $\beta$, and $\gamma$, respectively:

$$\frac{\gamma - \alpha}{330 - 235} = \frac{\beta - \alpha}{270 - 235} \tag{9}$$

From Equation (9), the loss data for M270-35A can be derived as:

$$\beta = \frac{12 \cdot \alpha + 7 \cdot \gamma}{19} \tag{10}$$

The iron losses of machine iron cores made of M270-35A in both CM and CR configurations are computed by linearly interpolating the losses of M235-35A and M330-35A for each respective configuration.

### 3.2.2.2. Method 2 (M2) – linearity of loss change

Equation (10) is employed to determine the specific iron losses of M270-35A at 77 K, for which the relative change values from RT to 77K are needed. Currently, the specific iron losses of M235-35A and M330-35A at RT and 77K [42], as well as the loss data of M270-35A at RT [39], are available. Those data are derived from measurements taken on materials specifically processed for the fabrication of electrical machines.

Rather than relying on the linear relationship among the losses of the three materials, M2 assumes that the relative change values from RT to 77K for the three materials also follow similar linearity to Equation (10). Consequently, Equation (10) can be employed to determine the relative change values from RT to 77 K for M270-35A, by using the relative change values of the specific iron losses of M235-35A and M330-35A.

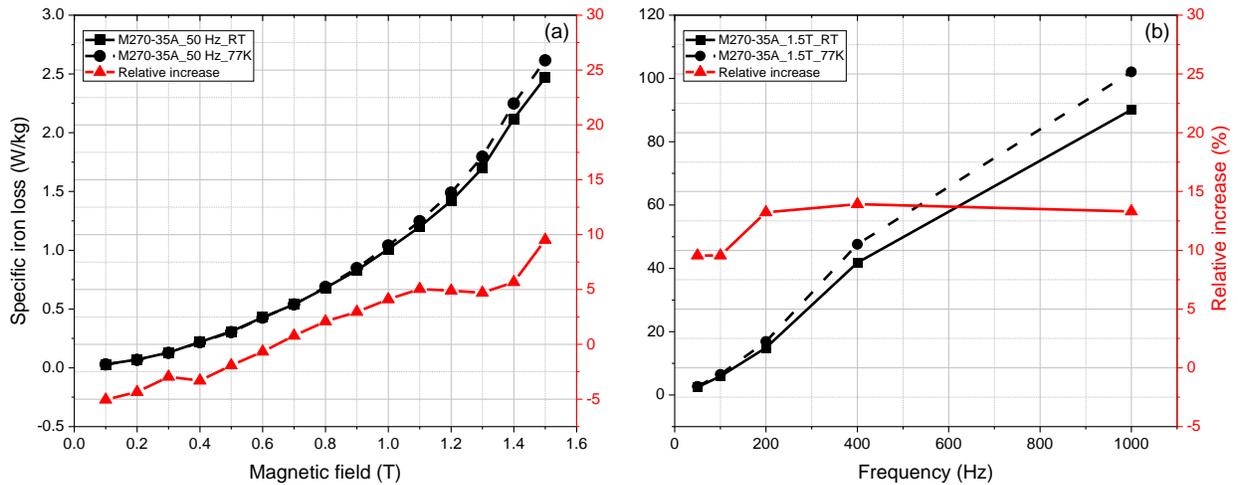

Figure 4. The specific iron loss of M270-35A for RT and 77 K as well as the relative increase from RT to 77 K at 50 Hz (b) 1.5 T. Data are obtained through linear interpolation M2.

The relative change values for the three materials are denoted as $\alpha_1$, $\beta_1$, and $\gamma_1$, respectively. By applying the obtained relative increase ratio $\beta_1$ to the specific iron losses of M270-35A at RT, obtained from the manufacturer's data sheet [44], the iron loss data at 77 K for M270-35A can be estimated. Two examples are given on 50 Hz and 1.5 T magnetic flux density, as shown in Figure 4 (a) and (b) respectively. The obtained results of the specific iron losses at 77 K are presented alongside the specific iron loss at RT, as well as the relative increase, in Figure 4.

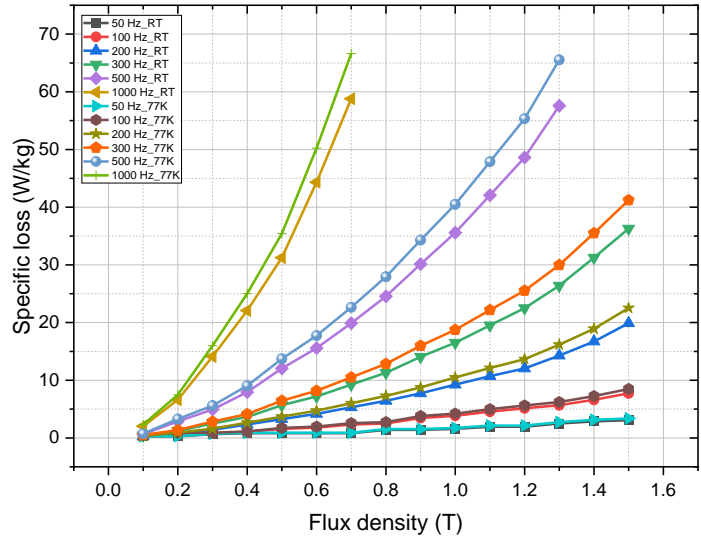

Figure 5. Specific iron loss at RT and 77 K obtained through curve fitting, with the external field varying from 0.1 T to 1.5 T and the frequency from 50 Hz to 1000 Hz.

By extending the calculation across all frequencies (50 - 1000 Hz) and flux densities (0.1 - 1.5T), specific iron losses for M270-35A can be derived, as depicted in Figure 5. It is evident that losses are elevated at 77 K compared to at RT and the disparities between the curves of RT and 77 K are more pronounced at higher magnetic flux densities and frequencies, consistent with the relative increases indicated in Figure 4.

The same approach as the one mentioned above for the calculation of iron losses is employed to calculate the electrical conductivity of M270-35A at 77K, which is listed in Table 4 along with the other characteristic parameters and fitted coefficients for the iron losses of M270-35A. As mentioned in Section 3.2.1, the loss coefficients for the Bertotti formula have been determined through Equation (7) and curve fitting with the software ANSYS Motor-CAD based on iron loss data in Figure 5. The computed coefficients for the losses are expressed in units of W·kg⁻¹, consistent with the specific loss power shown in Figure 5. For a direct and clearer comparison, the iron losses of machines with the iron steel grades M235-35A and M330-35A are also computed, with their iron loss coefficients taken from [42].

Table 4. Material parameters and loss coefficients for the iron loss calculation of machines made with M270-35A at RT and 77 K

| Parameter | Symbol | M270-35A |
|---|---|---|
| Exponent for magnetic flux density | $\alpha$ | 2 |
| Volumetric mass density [kg·m⁻³] | $\rho$ | 7650 |
| Lamination thickness [mm] | $d$ | 0.35 |
| Electrical conductivity at RT [S·m⁻¹] | $\sigma_{RT}$ | $1.92 \times 10^6$ |
| Electrical conductivity at 77 K [S·m⁻¹] | $\sigma_{77K}$ | $2.23 \times 10^6$ |
| Hysteresis loss coefficient at RT [W·kg⁻¹] | $K_{h\_RT}$ | $2.20 \times 10^{-2}$ |
| Eddy current loss coefficient at RT [W·kg⁻¹] | $K_{e\_RT}$ | $5.06 \times 10^{-5}$ |
| Excess loss coefficient at RT [W·kg⁻¹] | $K_{e\_RT}$ | $1.37 \times 10^{-3}$ |
| Hysteresis loss coefficient at 77 K [W·kg⁻¹] | $K_{h\_77K}$ | $2.32 \times 10^{-2}$ |
| Eddy current loss coefficient at 77 K [W·kg⁻¹] | $K_{e\_77K}$ | $5.87 \times 10^{-5}$ |
| Excess loss coefficient at 77 K [W·kg⁻¹] | $K_{e\_77K}$ | $1.65 \times 10^{-3}$ |

## 4. Results and discussion

This section presents the magnetisation results of the TFS, along with the copper losses in the stator magnetisation winding, and the iron losses at various magnitudes of the pulsed currents. The trapped field and flux, temperature rises, and magnetisation losses of the TFSs are demonstrated and analysed. Additionally, two cases are considered for iron losses during the magnetisation: one where the machine does not contain TFSs, and the other where one TFS is incorporated into the machine.

## 4.1. Magnetisation results

Figure 6 shows the magnetic field distribution within the simulated region, one eighth of the machine, and the distribution of current density and temperature within half of a stack with nine layers, when the pulsed current magnitude reaches 500 A as an example. The symmetry of the studied superconducting motor and HTS stack allows for deriving current and temperature distributions in the remaining half of the stack by mirroring the distributions observed in the current half. However, it is crucial to emphasise that the current distribution in the other half should be multiplied by a negative sign relative to that in the current half stack. With a self-field critical current being 391 A/cm at 77 K, the critical current density in the 1 μm superconducting layer is $J_{c0}$ = 4.31×10$^{10}$ A·m$^{-2}$. Specific instances are demonstrated: the pulse peak ($t$ = 2.25 ms) in Figure 6 (a), the pulse end ($t$ = 11 ms) in Figure 6(b), and the magnetisation end ($t$ = 30 s) in Figure 6(c). The depicted white arrows indicate the magnetisation field direction. At the pulse peak, the magnetic field within the air gap reaches its maximum intensity. In addition, the upper section of the iron yoke, situated directly above the stack, achieves its saturation field of 1.7 T, as shown in Figure 6(a) and maintains a field strength of approximately 0.9 T at the magnetisation end, as depicted in Figure 6(c).

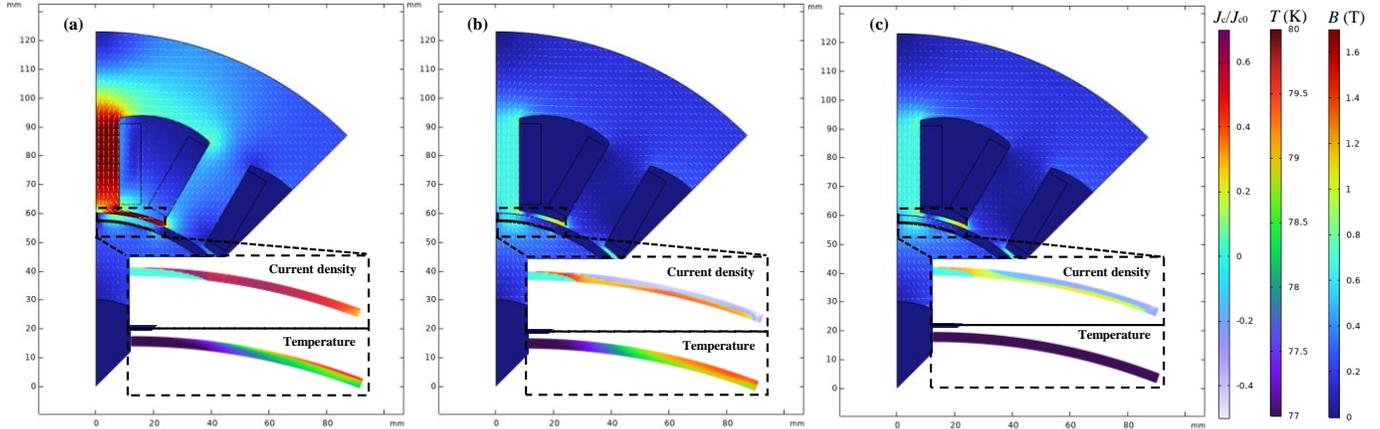

Figure 6. Magnetic flux density distribution within the one-eighth model and the distribution of current density and temperature within one half of a nine-layer stack at the (a) pulse peak (b) pulse end (c) end of the magnetisation, with a pulsed current magnitude of 500 A.

At the magnetisation end, the current distribution exhibits currents in two directions stacked from the top to the bottom of the TFS, indicating incomplete penetration of the magnetisation field (1.06 T) into the central stack region. This accounts for the observed M-shaped magnetic field distribution for the nine-layer stack in Figure 7. For full penetration, the current distribution would be a single macroscopic current circulating throughout the entire stack volume.

As an example, Figure 7 illustrates the trapped magnetic flux density at 1 mm above the stack surface, with a 500 A pulsed current and the tape layer number ranging from one to nine. For one to five layers, the HTS stack experiences complete penetration, showing a conical profile of trapped flux density. The trapped field at a distance of 1 mm above the central stack point rises from 0.15 T (single layer) to 0.63 T (five layers). However, for six layers and beyond, the magnetisation becomes partial, resulting in reduced trapped flux in the stack central region. Consequently, this leads to M-shaped profiles for six to nine layers, indicating an underutilisation of the stack's magnetisation capacity. This aligns with previous numerical [9] and experimental results [45].

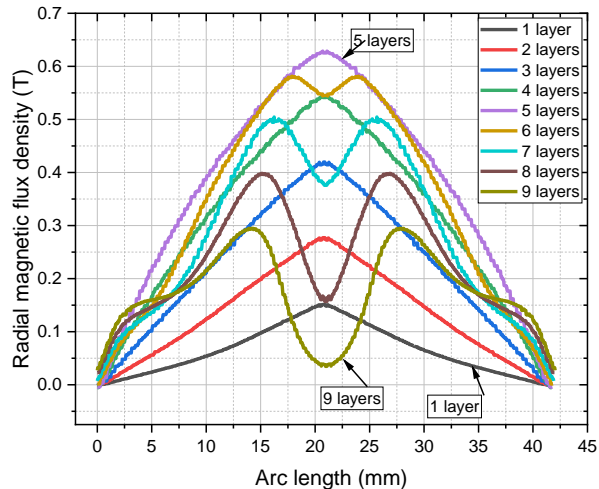

Figure 7. Waveforms of the trapped field extracted from 1mm above the stack surface with varying layer number for an applied current of 500 A.

In Figure 8, the average trapped field and the trapped flux at a distance of 1mm above the stack surface are presented, showing variations in layer number ranging from one to nine, along with applied pulsed current spanning from 250 A to 1 kA. Across all pulsed current levels, the average trapped field exhibits an initial ascent followed by a descent with increasing layer numbers, as depicted in Figure 8 (a). As the applied pulsed current escalates, the peak value undergoes an augmentation, rising from 0.28 T at four layers to 0.45 T at eight layers. The apex of the average trapped field distribution aligns with the highest layer number that can be fully penetrated. This alignment is evident in Figure 7, where the maximum layer number that can be completely penetrated is five for a pulsed current of 500 A while the average trapped field also peaks at five layers. To elaborate, the maximum fully magnetised layers correspond to three, five, seven, and eight for pulsed currents of 250 A, 500 A, 750 A, and 1000 A, respectively.

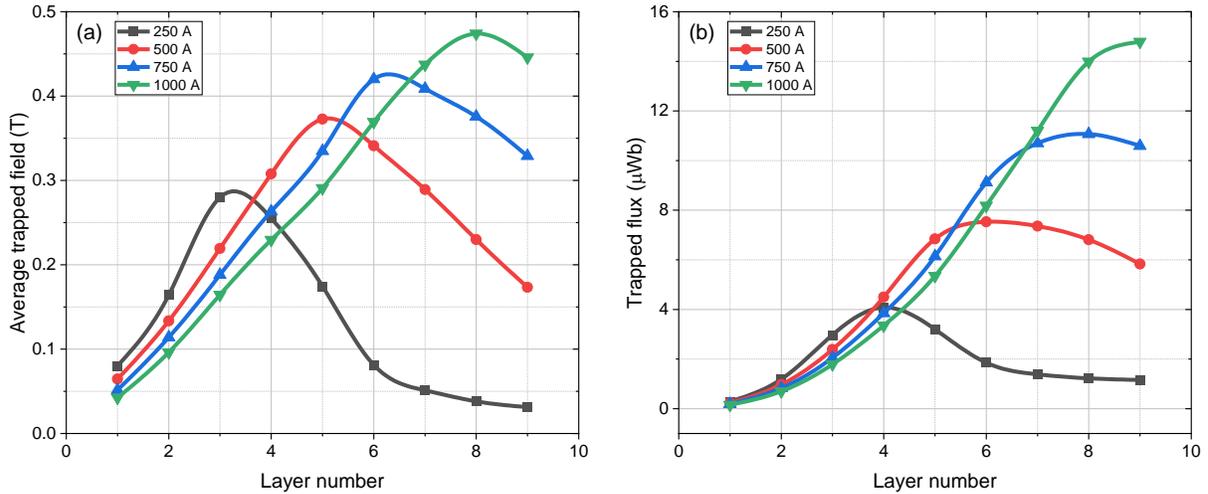

Figure 8. Variation of the average trapped field extracted from 1 mm above the stack surface with the stack layer number from one to nine and the pulsed current from 250 A to 1000 A.

The pattern of trapped flux resembles that of the average trapped field, as illustrated in Figure 8 (b). As the pulsed current amplitude increases, the maximum layer number that can be fully penetrated also rises, resulting in a higher trapped magnetic flux density and, consequently, a greater trapped magnetic flux in the TFS. However, the peak value of the trapped flux occurs at progressively higher layer numbers as the pulsed current magnitude grows. Specifically, for pulsed currents ranging from 250 A to 1000 A, the layer numbers corresponding to the maximum trapped flux are four, six, eight, and nine, respectively. In comparison, the number of layers for full penetration and peak trapped field are three, five, six, and eight, across various pulsed current amplitudes. Table 5 summarises the layer numbers for different characteristics at various pulsed current magnitudes, e.g., the full penetration, peak trapped field, and peak trapped flux.

Table 5. Layer number for full penetration, peak trapped field, and peak trapped flux of the HTS stack

| Pulsed current magnitude [A] | Layer number for full penetration | Layer number for peak trapped field | Layer number for peak trapped flux |
|---|---|---|---|
| 250 | 3 | 3 | 4 |
| 500 | 5 | 5 | 6 |
| 750 | 6 | 6 | 8 |
| 1000 | 8 | 8 | 9 |

Figure 9 demonstrates the temperature rise of TFSs as the number of layers increases at various pulsed current amplitudes during magnetisation. Two sets of temperature rise values are presented: the first set, depicted in Figure 9(a), represents the maximum temperature rise in the TFSs, which occurs typically at the pulse peak, while the second one, depicted in Figure 9(b), displays the peak values of the average temperature rise. The pulsed current amplitude of 1kA demonstrates the highest maximum temperature rise and average peak temperature rise across all layer numbers. Consequently, as depicted in Figure 8, the curves for trapped field and trapped flux exhibit reduced slopes with rising pulsed amplitudes prior to reaching their peak values.

The highest temperature is under 80 K, with a peak increase within 3 K. The magnetisation flux initially penetrates the stack from the top surface, given the placement of the magnetisation coils above the stack. Consequently, the temperature of the upper part of the stack increases before the lower part on both sides. The relative temperature rise in the stacks depends on layer numbers and magnetisation field levels but consistently remains below the critical temperature of YBCO in all simulation scenarios in Figure 8. The highest temperature rise observed is 9 K, occurring with one layer and a pulsed current of 1kA.

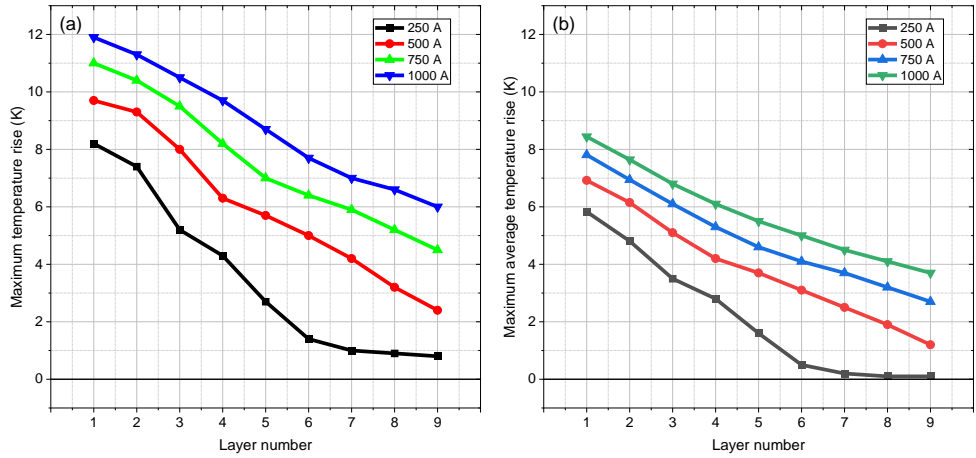

Figure 9. The temperature rises of TFSs with increasing number of layers at various pulsed current amplitudes during the magnetisation (a) the maximum temperature rises in the TFSs (b) the peak values of the average temperature rise.

The magnetisation power loss (W/m) generated during the dynamic process of PFM accounts for the majority of the loss during the magnetisation process and it can be calculated using

$$Q = \frac{1}{T}\int\limits_{0}^{T}\iint\limits_{S} E \cdot J \mathrm{d}S \mathrm{d}t \tag{11}$$

where $T$ is the pulse time duration, namely 11 ms in this paper.

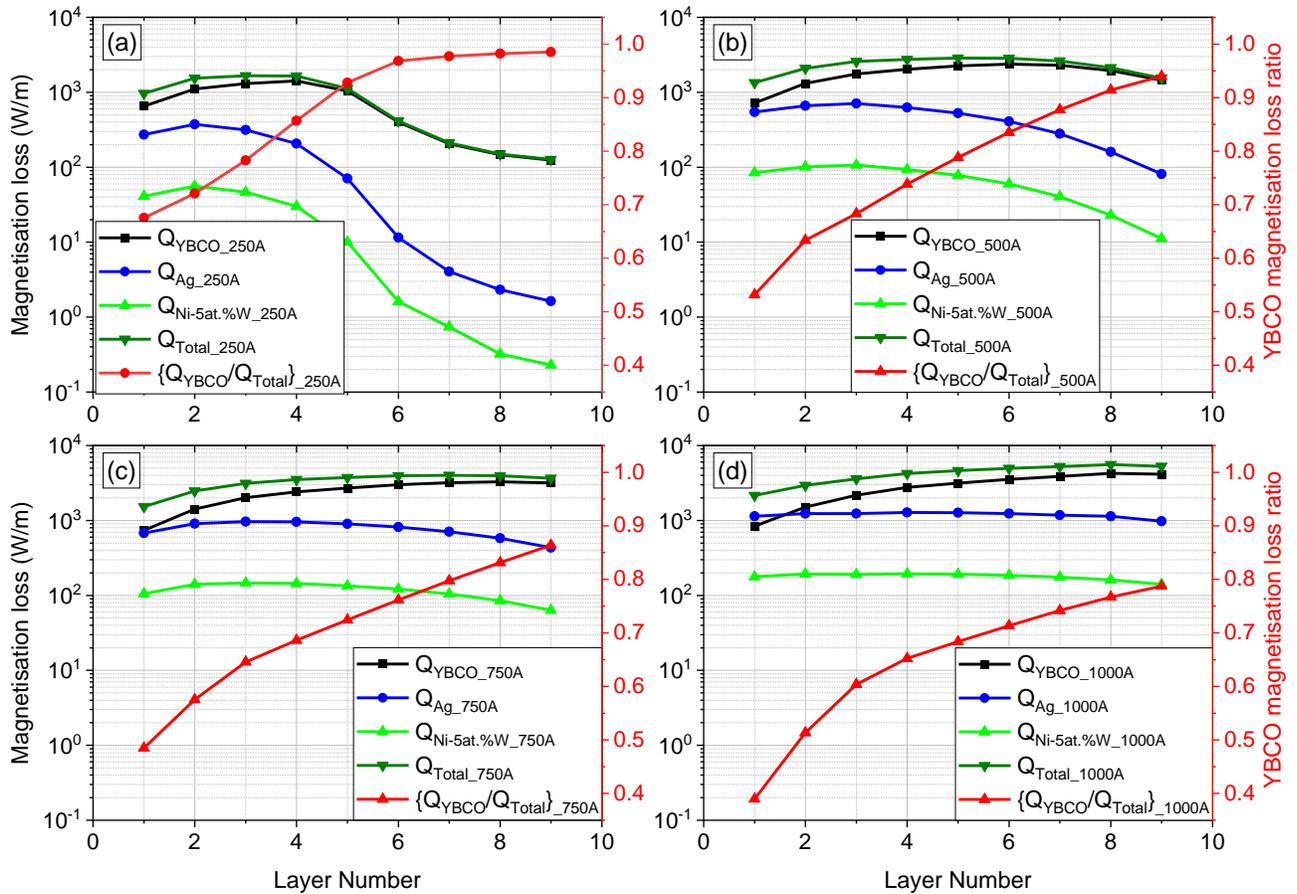

Figure 10. Magnetisation loss generation across various layers with the pulsed current ranging from 250 A to 1000 A and the layer number increasing from one to nine.

Figure 10 demonstrates the magnetisation losses within distinct layers, while the applied pulsed current spans from 250 A to 1000 A. Notably, the superconducting layers contribute most significantly to heat generation, while the silver layers also produce substantial heat due to eddy currents [46]. Interestingly, the heat generated within the substrate layers is considerably lower than that of the other layers, which is attributed to their considerably higher electrical resistivity relative to the other layers. Furthermore, with an increasing number of layers, the magnetisation loss ratio of the YBCO layers to the total magnetisation loss rises, while the ratio of heat produced in other layers to the total magnetisation loss diminishes.

Across all cases of pulsed current, heat generation in individual layers demonstrates an initial increase followed by a subsequent decrease as layer numbers increase. Consistent with the observations in Figure 8, the peak heat generation values in YBCO layers $Q_{YBCO}$ and the total heat $Q_{Tot}$ coincide with the maximum number of fully magnetised layers. The heat generation in the silver layer $Q_{Ag}$ and substrate layer $Q_{Ni-5at.\%W}$ declines with increasing layer numbers, indicating a reduction of the induced eddy currents within the metal layers. However, as the pulsed current increases, the decreasing trend becomes less pronounced due to the escalating heat generated by eddy currents.

### 4.2. Copper loss

The copper losses of the magnetisation winding at various current magnitudes are demonstrated in Figure 11. Figure 11 (a), (b), and (c) illustrate the specific copper loss power, the accumulated specific copper loss, and the accumulated copper loss within one pulse duration, respectively. The accumulated losses are determined by integrating the power loss over time. The specific accumulated losses are computed by dividing the accumulated losses by the weight of the stator copper windings. The weight of the copper windings employed for magnetisation can be found in Table 3. The copper loss power increases with higher input current magnitudes and it drops to zero at the pulse end due to the current magnitude dropping to zero.

The pattern of copper loss power corresponds closely to that of the magnetisation current shown in Figure 2 (a). This correlation arises because copper losses increase with the square of the magnetisation current. Moreover, copper losses at 77 K surpass those at RT due to the higher resistance of copper windings at 77 K compared to RT. The higher resistance is attributed to the increased electrical conductivity at 77 K. The accumulated copper loss during one single pulse ranges from 1.0 to 16.4 J at RT and from 1.2 to 18.7 J at 77 K. This amount of heat should be conducted away using appropriate cooling measures, making the CM configuration, where the stator is immersed in liquid nitrogen, a more suitable choice for the overall design.

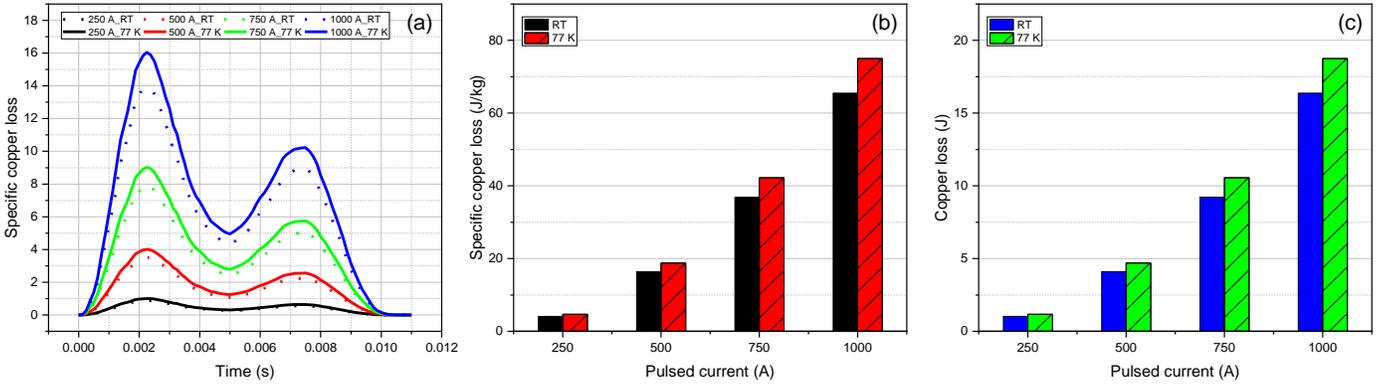

Figure 11. Copper losses in the magnetisation winding at RT and at 77 K for various magnetisation currents within a pulse duration (a) variation of the specific copper losses (b) accumulated specific copper losses (c) accumulated copper losses.

### 4.3. Iron loss

As previously discussed, the diamagnetic impact of HTS stacks on the iron losses was examined. Consequently, this section is divided into two parts: the first part analyses iron losses without the HTS stacks, while the second part examines the losses in the presence of the HTS stacks calculated using Method 2.

#### 4.3.1. Without HTS stacks

Figure 12 presents the total specific iron losses in the entire machine for different steel grades, (a) M235-35A (b) M270-35A (c) M330-35A, over one pulse duration in both CM and CR configurations across varying pulsed current amplitudes. The losses for M270-35A were calculated using Method 2. Notably, the iron losses are greater in the CM configuration compared to the CR configuration for both the stator and the rotor. This difference is due to the varying loss characteristics at RT and CT, as depicted in Figure 5. Consequently, the CM concept exhibits higher total specific iron losses than CR concepts.

When plotting the iron loss power over time during one magnetisation pulse, as illustrated in Figure 12, it is observed that the iron loss power exhibits a similar pattern to the copper loss power, mirroring the trend of the magnetisation current shown in Figure 2(a), in accordance with Equation (6). This similarity arises from the fact that iron losses are dependent on both the frequency and magnitude of the magnetisation magnetic field. Given that the frequency components remain constant, the iron loss power solely depends on the magnetic field, following exponents in the range of 1.5 to 2, while the magnetic field in turn correlates with the magnetisation current. Additionally, no significant losses occur after the pulse ceases, as the magnetic field in the iron yokes can be considered negligibly small.

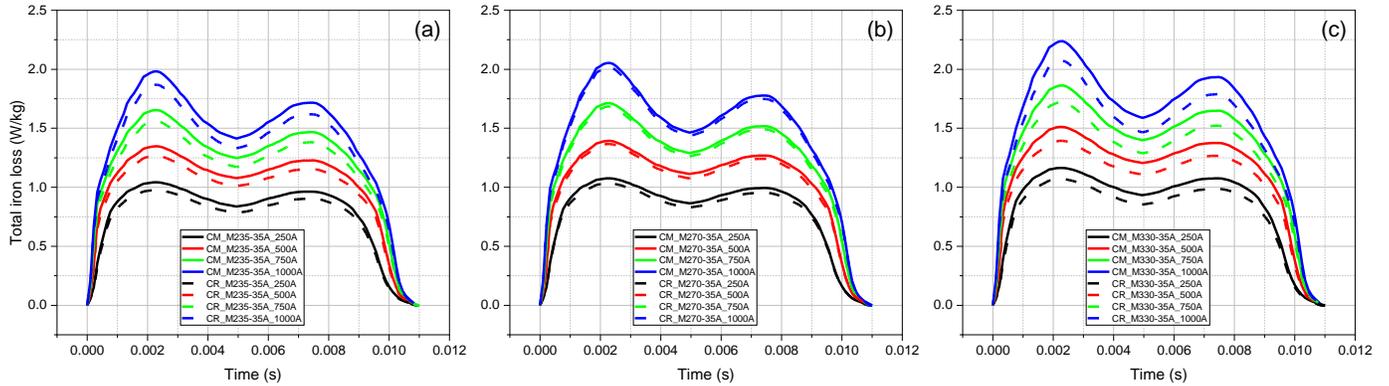

Figure 12. The specific total iron loss of the machine for the CM and CR configurations across various pulsed current magnitudes within one pulse duration for different materials (a) M235-35A (b) M270-35A (c) M330-35A. The data for M270-35A was computed using Method 2.

Figure 13 presents the accumulated losses (integrated over one pulse duration) of the machine iron cores - including total loss, stator loss, and rotor loss - in the CM [see Figure 13 (a)] and CR [see Figure 13 (b)] configurations, respectively. Both sub-figures illustrate loss results for cases where the machine is assumed to be made of M235-35A, M270-35A, and M330-35A. The iron losses of the electrical machine for M270-35A were calculated using Method 1 and Method 2, showing minimal differences in the rotor iron losses within each configuration (CM and CR). Similarly, the stator losses and, consequently, the total losses calculated by both methods remain consistent across both configurations. Notably, the iron losses for M270-35A, as determined by both methods, lie between those of M235-35A and M330-35A. This observation aligns with the anticipated relationship between iron losses and steel grades, confirming the expected trend.

In the CM concept, the total specific loss of the machine iron cores averages from 9 to 16 mJ/kg for all three steel grades across the pulsed amplitude. In contrast, the iron loss in the CR configuration spans from 8 to 15 mJ/kg. It is worth noting that the rotor iron losses are consistently lower compared to the stator iron losses in both the CM and CR configurations across all the pulsed current magnitudes, as depicted in Figure 13. This is attributed to the smaller magnetic field present in the rotor iron compared to that in the stator iron yoke during magnetisation, in accordance with Equation (6). This field discrepancy arises due to the increased reluctance resulting from the airgap.

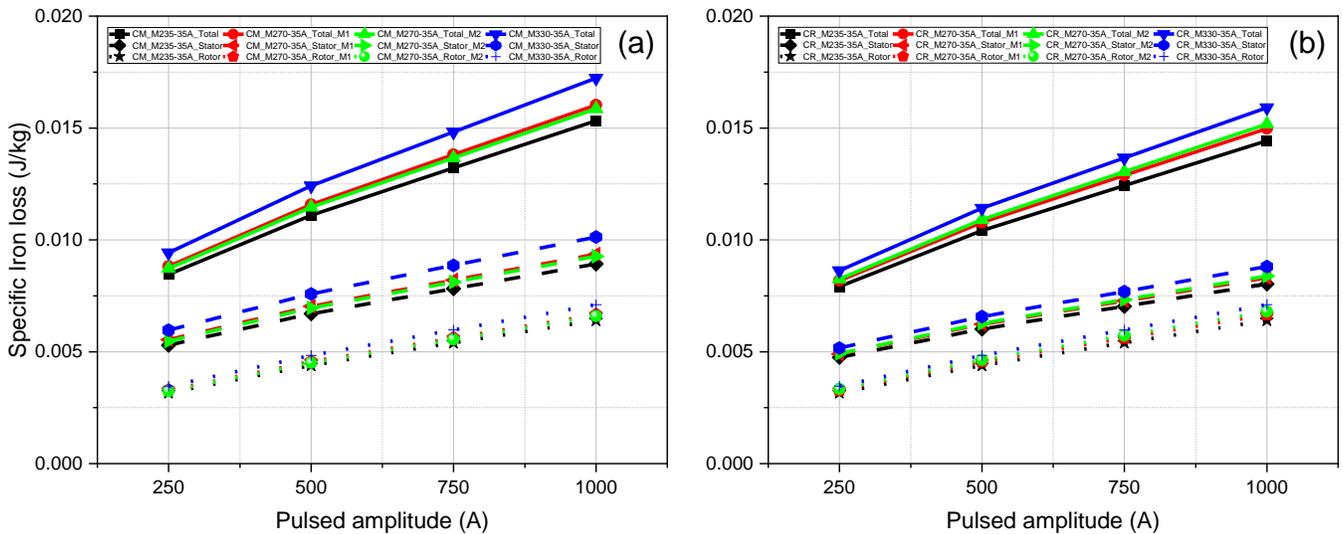

Figure 13. Iron losses - total, stator, and rotor iron losses - for the cases where the machine is assumed to be made of M235-35A, M270-35A, and M330-35A in the (a) CM and (b) CR configurations, respectively. The iron losses for M270-35A were calculated using Method 1 (M1) and Method 2 (M2).

In the CM configuration, the total losses, stator losses, and rotor losses of M270-35A calculated using Method 1 are higher than those calculated by Method 2. In contrast, the opposite trend is observed in the CR configuration. This discrepancy could potentially be attributed to the different calculation methods used: Method 1 relies on interpolation, whereas Method 2 utilises various coefficients to account for different loss components and temperatures. These methodological differences might affect the accuracy and consistency of loss calculations across different configurations. Despite this, the overall difference in losses between Method 1 and Method 2 for M270-35A remains below 3%, which is within an acceptable range.

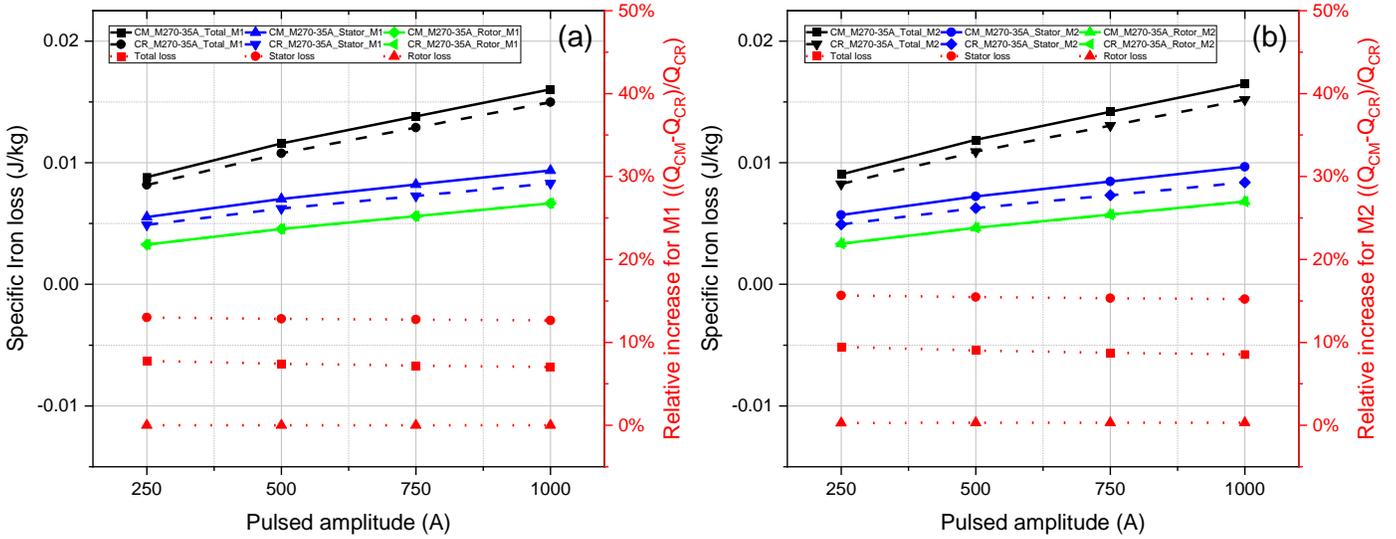

Figure 14. Iron losses - total, stator, and rotor iron losses - calculated with (a) Method 1 and (b) Method 2. The relative increase from the CR configuration to the CM configuration is also depicted.

Figure 14 depicts the total losses, stator losses, and rotor losses of M270-35A for the CM and CR configurations calculated using Method 1 [Figure 14(a)] and Method 2 [Figure 14(b)], respectively. In addition, the relative difference of these losses between the CM and CR configurations is presented in both figures. Rotor losses for M270-35A calculated by both Method 1 and Method 2 are almost equal in both CM and CR configurations. This is because the rotor operates at the same temperature in both configurations, leading to a negligible difference in relative increase for rotor losses across the pulsed current amplitudes.

The results of the stator and total losses indicate that both Method 1 and Method 2 yield comparable outcomes for both CM and CR configurations. For M270-35A, both methods show similar discrepancies in stator losses and total losses between the CM and CR configurations. Specifically, there is an average difference of approximately 14% in the stator losses between the CM and CR configurations, reflecting the variation in losses between RT and 77K for M270-35A. The average difference in total losses between Method 1 and Method 2 is around 9%, which can be primarily attributed to variations in the stator losses.

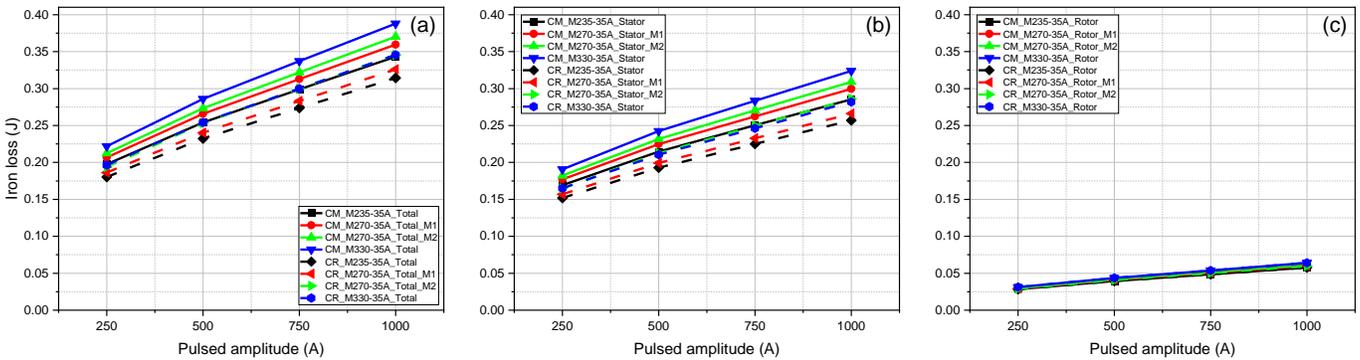

Figure 15. Accumulated iron losses during the PFM in the CM and CR configurations with varying pulsed current amplitude in the (a) entire motor, (b) stator, and (c) rotor, respectively.

To provide a rough estimation of the order of magnitudes for the iron losses, Figure 15 illustrates the accumulated total losses [Figure 15 (a)], stator losses [Figure 15 (b)], and rotor losses [Figure 15 (c)] in both the CM and CR configurations across various pulsed current amplitudes for different materials. Consistent with the trends observed in Figure 13 and Figure 14, there is a minimal discrepancy in rotor losses, whereas stator losses and total losses exhibit relatively larger differences across various configurations and materials. Overall, rotor losses range from 0.03 J to 0.07 J across varying pulsed amplitudes and configurations. On average, stator losses span from 0.18 J to 0.29 J with an error margin of ±0.025J, while the total losses range from 0.21 J to 0.35 J with an error margin of ±0.03J.

### 4.3.2. With HTS stacks using Method 2

When taking the HTS stacks into consideration, the presence of HTS CCs on the rotor surface acts as a shield on the rotor iron core, further reducing iron losses in the rotor core. Consequently, rotor iron losses will make up even smaller percentages in the total losses when taking the HTS stacks into consideration.

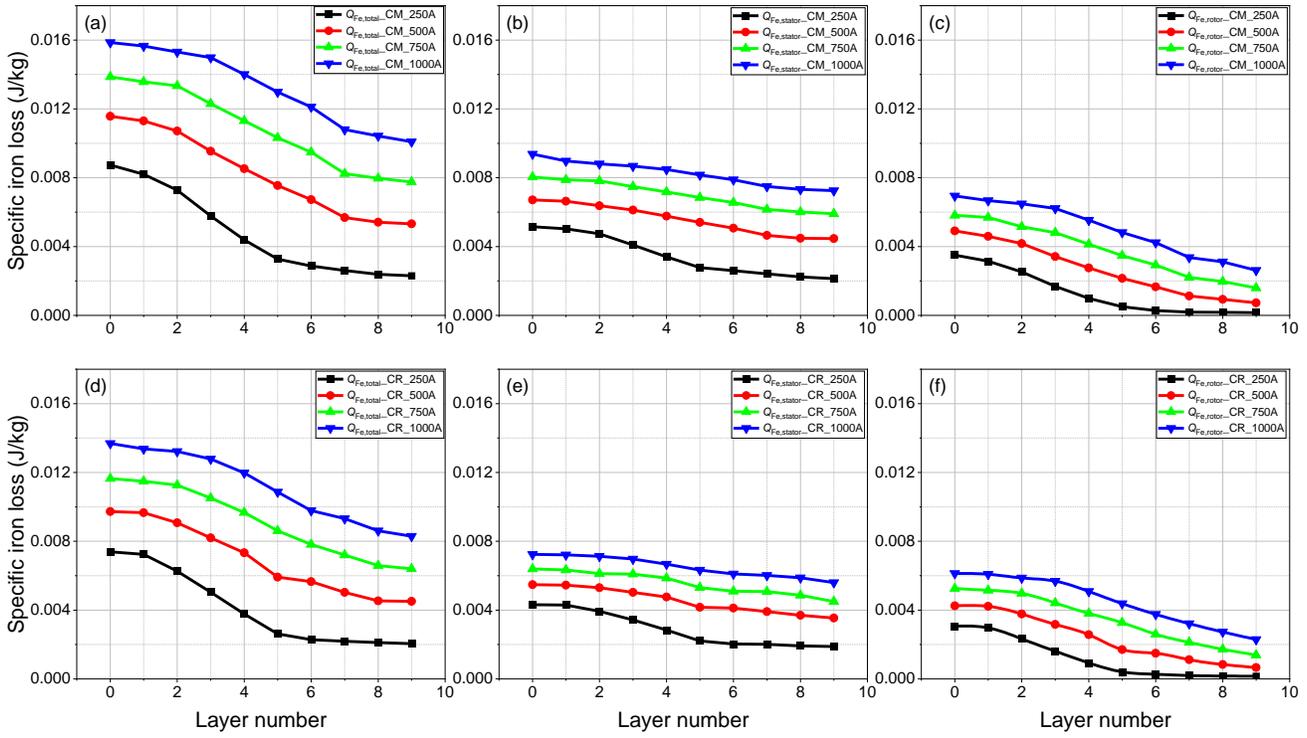

Figure 16. Specific iron losses during the PFM in the entire motor, stator, and rotor with varying pulsed current amplitude and varying tape layer number in the CM ((a)-(c)) and CR ((d)-(f)) concepts, respectively.

Figure 16 displays the specific iron losses during the PFM in the entire motor, stator, and rotor calculated using Method 2 for different tape layer numbers varying from zero to nine. As the pulsed current magnitude rises, the iron losses increase across layer numbers. The maximum specific iron losses of the entire machine are observed at the highest pulsed current amplitude without the presence of HTS stacks, ranging from 9 mJ/kg to 16 mJ/kg. This corresponds to an accumulated loss of approximately 0.23 J to 0.35 J. Notably, for a given pulsed current amplitude, the iron losses decrease with the rising layer number. These trends are visible in all the loss types for both configurations.

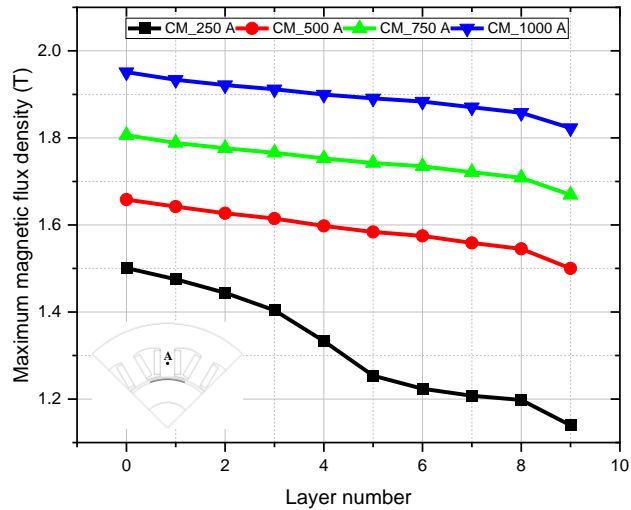

Figure 17. The maximum magnetic flux density in the middle point of the iron part between the two magnetisation coils.

As observed in Equations (6) and (8), the iron losses during the PFM are solely dependent on the strength of the magnetic flux density, as other factors including the frequency remain constant. Therefore, the trend of iron losses reflects the changing trend of the magnetic field intensity in the stator and rotor iron yokes. Figure 17 illustrates the maximum magnetic flux densities within the pulse at the midpoint of the iron section between the two magnetisation coils (probe point A in the inset of Figure 17) in the CM configuration, considering various layer numbers and pulsed current amplitudes. The highest magnetic field, measured at the maximum pulsed current amplitude without the presence of HTS stacks, reaches 1.95 T. Conversely, the lowest magnetic field, observed at the minimum pulsed current amplitude with a nine-layer stack, is 1.13 T. It can be observed that the magnetic flux density increases with the pulsed current amplitude. Notably, the magnetic flux density decreases as the number of layers increases, due to the enhanced magnetic shielding capabilities of the HTS stack with more layers.

### 4.3.3. Discussion on the iron loss

In this subsection, the three types of losses in iron yokes - hysteresis loss, eddy current loss, and excess loss – for the studied machine using M270-35A steel in various configurations (CM and CR) will be presented and analysed. Figure 18 demonstrates these losses in the entire machine across different pulsed current magnitudes for M270-35A in the CM [Figure 18 (a)], and CR [Figure 18 (b)] configurations, respectively.

As indicated in Equation (6), all three components of iron loss increase with the magnetic field magnitudes in the iron cores, which are proportionally influenced by pulsed current amplitudes. This relationship is evident from the increasing values of all three types of losses in both the CM and CR configurations across varying pulsed current amplitudes, as illustrated in Figure 18.

In both configurations, hysteresis losses account for the majority among the three types of iron losses across the pulsed current amplitudes, followed by excess losses, with eddy current losses being the smallest. This trend aligns with the findings in [42], where hysteresis losses comprise a larger percentage compared to excess and eddy current losses during motor operations.

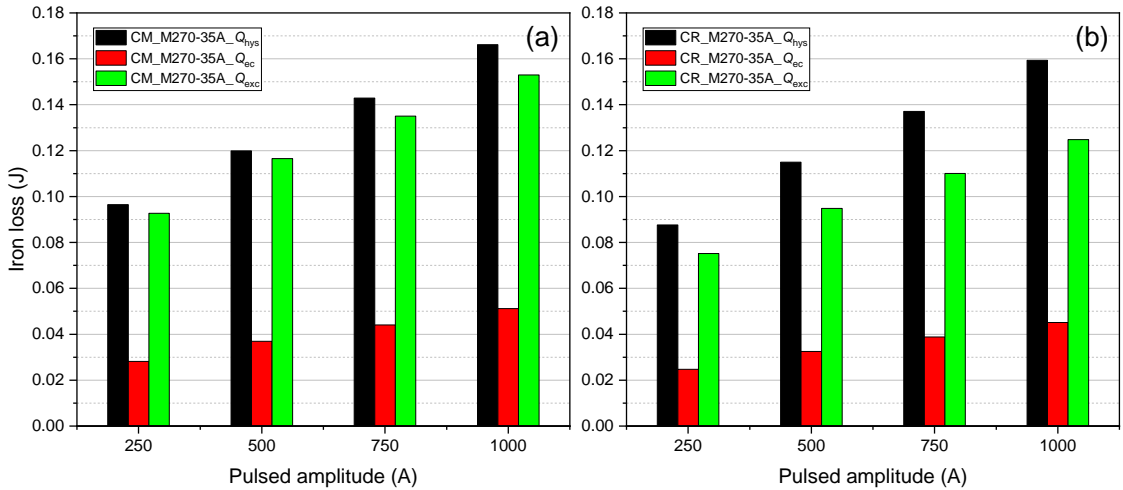

Figure 18. Comparison of the hysteresis loss ($Q_{hys}$), eddy current loss ($Q_{ec}$), and excess loss ($Q_{exc}$) for the M270-35A machine without HTS stacks in the (a) CM and (b) CR configuration.

By comparing the losses in the CM and CR concepts, it can be inferred that temperature significantly influences the losses. As temperature decreases, the relative permeability of the iron cores increases, which is represented by an increase in the loss coefficient $K_h$, leading to higher hysteresis losses. This phenomenon results in greater hysteresis losses in the CM configuration compared to the CR configuration, primarily due to the lower operating temperature of the stator in the CM setup.

The same trend is observed for eddy current losses. According to Equation (7), the coefficient of eddy current losses depends on the thickness of the lamination and the electrical conductivity of the steel. Since all three materials have the same lamination thickness, the determining factor for the loss coefficient $K_c$ is the electrical conductivity. As shown in Table 4, the electrical conductivity of the iron cores increases with decreasing temperature, resulting in higher eddy current losses. Consequently, the eddy current losses are greater in the CM configuration than in the CR configuration, again due to the lower operating temperature in the CM setups.

## 5. Conclusion

This research is a follow-up to our last publication, which is based on a validated modelling methodology. This paper systematically explored the application of PFM to an HTS tape stack within a practical superconducting motor with cryocooled iron cores. The investigation involved varying the stack's layer number from one to nine and adjusting the pulsed current amplitude from 250 to 1000 A, leading to the applied field for magnetisation increasing from 0.87 T to 1.36 T in the air gap. With increasing pulsed current, the peak value of the average trapped field increases from 0.28 T at 250 A to 0.45 T at 1000 A, while the peak trapped flux rises from 4.08 μWb at 250A to 14.79 μWb at 1000A. Magnetisation losses have been calculated and analysed for various pulsed current amplitudes across each layer of the HTS tapes. An increase in the magnetisation loss ratio can be observed in the YBCO layer as the layer number increases, regardless of the pulsed current amplitudes.

In addition, copper and iron losses have been calculated during the PFM process. The accumulated copper loss during one single pulse ranges from 1.0 to 16.4 J across various pulsed current amplitudes at RT and from 1.2 to 18.7 J at 77 K. For iron loss calculations, two different concepts, CM and CR, have been considered and analysed in scenarios with and without HTS stacks. A linear relationship has been proposed to estimate the iron losses of M270-35A by utilising the loss characteristics and coefficients of two other materials with different steel grades, M235-35A and M330-35A. Two calculation methods were developed: the first involves direct loss interpolation, while the second uses linear interpolation of loss coefficients. The second method was used to verify the results obtained from the first method. The various components of the iron losses have been demonstrated and compared.

Key conclusions can be drawn as follows:

(1) An initial rise followed by a subsequent decline has been observed in the average trapped field as the number of layers increases. Notably, the peak aligns with the maximum layer number that allows for complete penetration. For example, with a progressive escalation of the applied pulsed current, the peak value of the average trapped field demonstrates an increase, ascending from 0.28 T for four layers at 250 A to 0.45 T in the case of eight layers at 1000 A.

(2) When comparing the two proposed configurations (CM and CR), the system energy efficiency can be enhanced by the CR concept due to reduced losses within the motor's active materials. Specifically, copper losses are lower due to reduced electrical resistivity, and iron losses are minimised due to lower relative permeability and electrical conductivity. In addition, the CR concept requires less cooling power and has a shorter cool-down duration for the stator, as it operates at RT. However, this approach necessitates a design that ensures effective sealing to prevent coolant leakage from the air gap and maintain thermal isolation between the motor's stator and rotor. Before these issues can be solved, the CM configuration is more practical, as it can effectively dissipate the heat generated during the magnetisation.

(3) Across the varying pulsed current amplitudes and configurations, the total iron losses range from 0.21 J to 0.35 J on average. The stator iron losses range from 0.18 J to 0.29 J, while the rotor iron losses span from 0.03 J to 0.06 J. Among the three types of iron losses, hysteresis losses are the largest across the pulsed current amplitudes, followed by excess losses, with eddy current losses being the smallest. Iron losses in the CM configuration are higher compared to the CR configuration, primarily because iron losses increase with decreasing temperatures.

(4) In surface-mounted topology, TFSs reduce iron losses due to their shielding effect, which diminishes the magnetic field in the motor iron cores. For a given magnetisation current, the shielding effect becomes more pronounced as the number of HTS tapes increases. In the studied magnetisation process, the highest magnetic field is measured at the maximum pulsed current amplitude without the presence of HTS stacks, reaching 1.95 T. Conversely, the lowest magnetic field, observed at the minimum pulsed current amplitude with a nine-layer stack, is 1.13 T.

Overall, this study has presented a comprehensive approach that serves as an important tool for enhancing the efficiency and performance of HTS motors through the optimisation of the magnetisation process and the minimisation of associated losses. This approach not only aims to determine the optimal layer number for achieving the maximum trapped field in HTS motors but also considers the overall losses in both the superconducting and non-superconducting parts of the machine. By addressing these critical aspects, the study provides valuable insights and methodologies that can significantly improve the design and operation of HTS motors.